\documentclass[12pt]{iopart}
\usepackage{hyperref}

\usepackage{orcidlink}

\usepackage{subcaption}

\usepackage{footmisc}

\begin{document}

\title[A Hands-on Experience with a Novel Scintillation Detector for Particle Physics]{A Hands-on Experience with a Novel Scintillation Detector for Particle Physics}

\author{Anja~Bitar$^1$, Andrea~Brogna$^1$, Fabian~Piermaier$^1$, Steffen~Schönfelder$^1$,
Stefan~Schoppmann\,\orcidlink{0000-0002-7208-0578}$^1$ and Quirin~Weitzel$^1$}
\address{$^1$ Detektorlabor, Exzellenzcluster~PRISMA$^+$, Johannes~Gutenberg-Universität~Mainz, 55128~Mainz, Germany}
\ead{stefan.schoppmann@uni-mainz.de}

\begin{abstract}
Particle physics, when taught in the classroom or lecture theatre, suffers from a lack of practical experience by students. Thus, we describe the construction of a fully working small particle physics detector using state of the art detector technology for demonstration in educational context.
Most of our setup can be constructed with relatively moderate effort, given that a home-level 3D-printer, a photosensor and readout electronics (at least an oscilloscope) are available.
\end{abstract}

%
\vspace{2pc}


%
%

\section{Introduction}
Particle physics is a sub-discipline of physics which is nowadays an integral part of the curriculum of undergraduate and secondary studies~\cite{Lehrplan_RLP,Lehrplan_JGU}.
Despite its relevance, teachers at all levels struggle to provide practical experience in particle physics, due to the complexity of measurement instruments and hazards of the typically involved ionising radiation.
If any, practical experience is limited to historic detection devices with little to no relevance in present research.
As a consequence, teaching particle physics is reduced to largely non-engaging activities where the teacher has the active part.

In this article, we present a possibility for a hands-on experience on novel technology for particle physics detectors.
This novel technology uses scintillation light and is introduced in \autoref{sec:scintillationdetectors}.
In \autoref{sec:construction}, we describe the construction process of our setup, while we describe the measurements one can perform with it in \autoref{sec:measurements}.
A conclusion is given in \autoref{sec:conclusion}.

\section{Scintillation Detectors in Particle Physics}
\label{sec:scintillationdetectors}
Scintillation detectors work by conversion of kinetic energy of an ionising particle into visible light, called scintillation light~\cite{Birks_1964}.
In our device, we produce scintillation light in aromatic hydrogen-carbon molecules.
An ionising particle, when striking such an aromatic molecule, ionises it by ejecting a bound electron from the molecule or it excites the molecule by promoting a bound electron to a higher energy level.
During the recombination of the electrically positive molecular ion with the negative electron or during the de-excitation of the excited molecule, binding energy is released in the form of ultra-violet (UV) photons.
This light can be detected by very sensitive photosensors, e.g.~silicon photomultipliers (SiPMs), when the photons undergo photoelectric effect in the photosensors.
In the photosensors, the initial charge carriers created by the photoelectric effect are triggering avalanches of secondary charge carriers, because they are accelerated inside a strong electric field.
Ultimately, the discharge becomes large enough to be picked up as electrical current at the outside of the photosensors.

\begin{figure}[t]
    \centering
        \includegraphics[width=\textwidth]{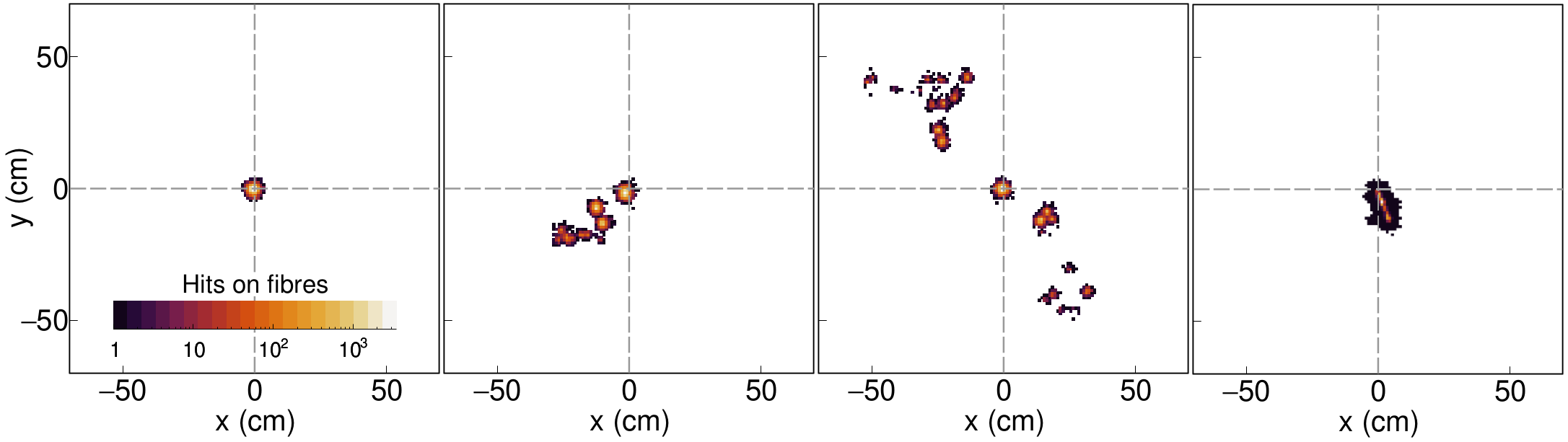}
        \caption[Simulation of an electron, a gamma-ray, a positron, and a muon]{Computer simulation of hits of optical photons (from left to right) of an electron, a gamma-ray, a positron, and a muon, all with the same energy of 2.8\,MeV\footnotemark~in an opaque scintillator medium traversed by optical fibres. The fibres are arranged along the $z$\mbox{-}direction in a triagonal lattice of 1\,cm pitch. While the electron shows a single energy deposition due to ionisation, the gamma-ray deposits energy at several vertices through multiple Compton-scattering off electrons~\cite{Compton_1923}. The positron combines the patterns of an electron and two gamma-rays, which are created during the annihilation of the positron. The muon produces light in a track. Muons with more energy, such as cosmic muons, would produce much longer tracks. In a transparent scintillator medium, the individual energy depositions are washed out into a large ball of light and particles cannot be distinguished. Figures are adapted from Reference~\cite{NuDoubt} with permission.}
    \label{fig:opaque_topology}
\end{figure}
\footnotetext{An energy of 2.8\,MeV $\approx 4.5 \cdot 10^{-13}$\,J.}

While the technology of scintillation detectors has been used in particle, nuclear, and medical physics for decades~\cite{Shwartz_2017}, in recent years some new ideas around aromatic liquid scintillators emerged~\cite{Schoppmann_2023}.
One of these ideas is the introduction of opacity into the otherwise transparent scintillator medium put forward by the LiquidO collaboration~\cite{LiquidO_2019,LiquidO_2024_LIME,LiquidO_2024}.
This can e.g.~be achieved by the introduction of wax.
The size of the wax structures and thereby their opacity can be controlled by temperature~\cite{Buck_2019}.
In an opaque scintillator, any light created by an ionising particle is confined to a small region of space, because the optical photons scatter multiple times and cannot travel in a straight line.
The light is then picked up locally by optical fibres that run through the opaque medium and which transport the light to photosensors at the outside of the detector.

As can be seen in \autoref{fig:opaque_topology}, this new technology allows to distinguish for the first time between electrons (beta-minus-radiation), gamma-radiation, and positrons (beta-plus-radiation) in a scintillation detector by their pattern of energy depositions, because it confines the scintillation light to the spot where the energy deposition is~\cite{LiquidO_2019}.
In a transparent scintillator, the structures seen in \autoref{fig:opaque_topology} are fully blurred, because the scintillation light is not confined, but travels through the entire transparent scintillator.
Thus, the invention of opaque scintillator allows to distinguish the individual particles reliably.

\section{Construction of a Particle Detector Prototype}
\label{sec:construction}
In this section, we describe the design and preparation of all components for a small opaque scintillation detector, as well as its final assembly.
Our setup simplifies the original setup used in Reference~\cite{LiquidO_2019}.
General requirements are liquid tightness, especially for oily substances, and light tightness if the detector is operated in a room with daylight.
The detector is designed such that measurements can be performed with both a pulsed light source (see \autoref{sec:measurements:temperature}) or with cosmic-ray muons (see \autoref{sec:measurements:muons}).
More technical information like CAD drawings and STL data links can be found in~\ref{sec:appendix}.
The readout system of the detector is described in \autoref{sec:CAENkit}.

\subsection{Detector Design}
\label{sec:design}
\begin{figure}[tb]
    \centering
        \includegraphics[scale=0.35]{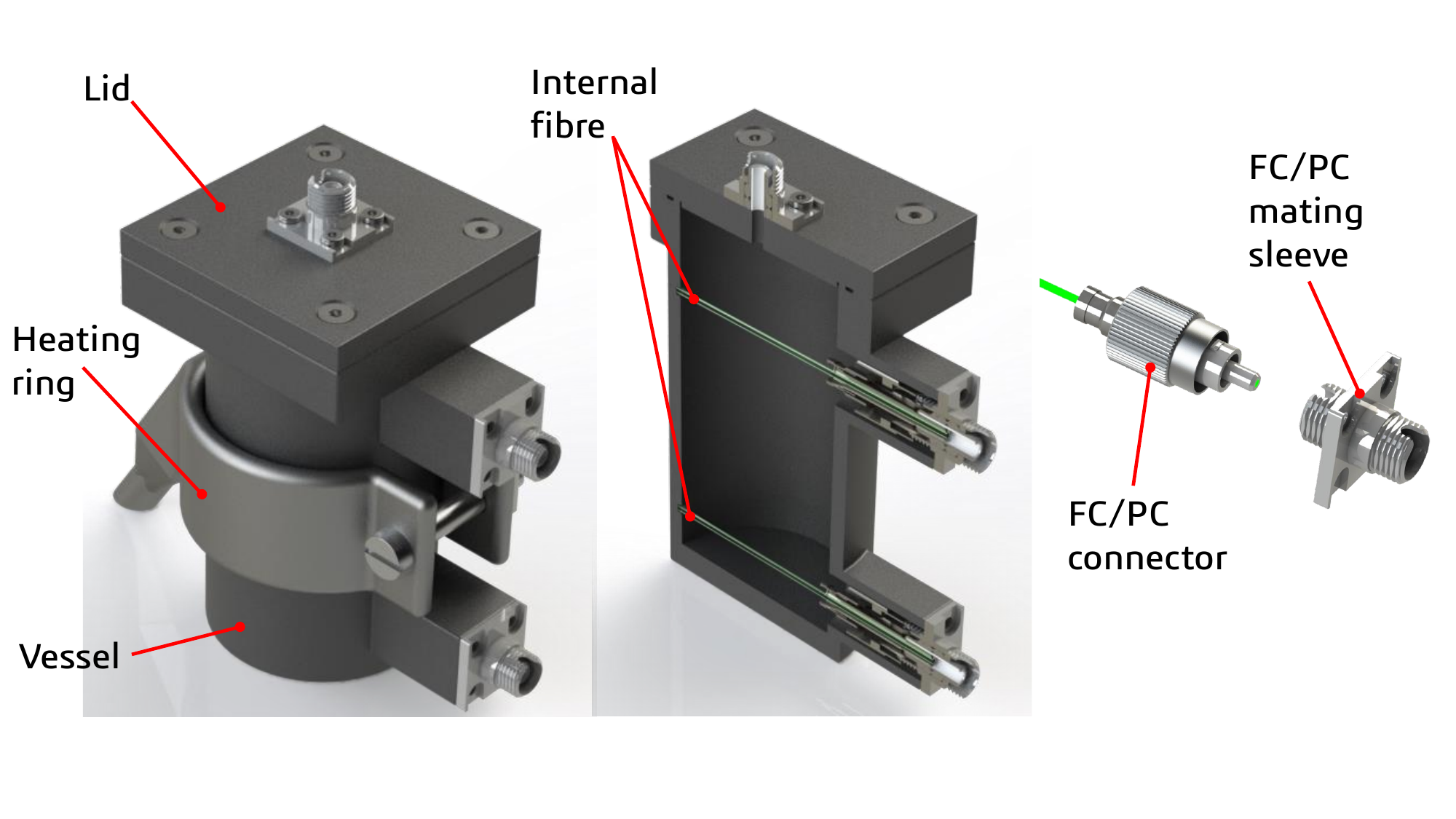}
        \caption{Artistic rendering of the detector (from left to right): with heating ring, cut in half, detail view of an internal fibre with FC/PC connection.}
    \label{fig:schnitt}
\end{figure}

\autoref{fig:schnitt} shows an artistic rendering of the detector.
The connector shown on the top-side allows for coupling of a light source\footnote{We use ADAFC2 connectors from Thorlabs, Inc. \url{https://www.thorlabs.com}}.
Inside the vessel are two wavelength shifting fibres, type Kuraray\footnote{Kuraray Co., Ltd. \url{https://www.kuraray.com}} Y-11 (200) M 1.00 mm D, which end in FC/PC connectors.
These are used for connection with SiPMs via optical patch cables. 
Utilising fibres with a core diameter of 1\,mm and FC/PC connectors ensures compatibility with the optical fibres of the CAEN kit, which will be introduced in \autoref{sec:CAENkit}.
Using only two internal fibres is the simplest design to show the opacity change of the scintillator when warmed up.
This could be expanded by introducing more fibres into the vessel.

The general shape of the detector as it is shown in \autoref{fig:schnitt} is mainly dictated by three boundary conditions: the dimensions of the FC/PC connectors and mating sleeves, the dimensions of the existing heating rings, and the optical properties of the opaque scintillator.
The main limiting geometrical factor is the heating ring. If other heating devices are used, e.g.~flat Peltier elements, the geometry of the vessel should be changed accordingly.
The detector consists of two parts: a lid, which carries the connector for the light source, and a vessel which houses the two internal fibres and the liquid scintillator.
The lid is attached to the vessel with four screws, making it detachable for easier cleaning if the scintillator needs to be changed. Brass thread inserts are used to ensure durable threads inside the 3D printed parts (see \autoref{sec:3Dprinting}). Thread inserts are also utilised to attach the FC/PC mating sleeve to the lid, but due to the lack of space for said inserts, the mating sleeves on the vessel have to be glued in.
The vessel has two small blind holes opposite of, but on axis to, the connectors in order to fix the position of the internal fibres.

The final result is a compact detector cell, about $9 \times 5 \times 7.5$\,cm\textsuperscript{3}, with a volume of around $50$\,ml of liquid scintillator, and CAEN kit compatibility.
Further technical details can be found in~\ref{sec:appendix}.

\subsection{3D Printing and Post-processing}
\label{sec:3Dprinting}
Since it is a flexible and convenient process, we employ 3D printing for manufacturing.
We have tested two different 3D printing techniques: Fused Deposition Modelling (FDM), also known as Fused Filament Fabrication (FFF), and Masked Stereolithogrophy (MSLA). For this work we use the model X7 from Markforged\footnote{Markforged, Inc. \url{https://markforged.com}}, and the model Mars 4 Ultra from ELEGOO\footnote{ELEGOO, Inc. \url{https://elegoo.com}}, respectively.
Both offer the possibility to print parts in different colours which can be used to influence whether the inner walls are diffusely reflective or more absorbent, but one has to keep in mind that the colour can also influence the light tightness. Following the design described in~\autoref{sec:design}, we have printed the detector in two parts.
To ensure liquid and light tightness, without having to make the wall thickness of the parts too thick, we coat the parts in a separate post-processing step.

\subsection*{FDM Printing}
We recommend using filament materials that are capable of maintaining structural integrity at temperatures up to 50\mbox{--}70\,\textcelsius~for prolonged periods of time.
Because of that we recommend e.g.~ABS, PETG, or other durable plastics.
Here, we have utilised PA 6 CF (Polyamide 6 with chopped carbon fibres) for its mechanical properties and since it is a standard material for our printer.
PLA is not recommended due to its low heat resistance.
With the Markforged X7, the printing time was about 6.5~hours.
After printing, both the vessel and the lid are sealed with a UV-curing resin on the outer surface.
For this, the resin from the MSLA printer has been used.
We have applied the resin onto the outer surfaces using a brush, followed by exposure to UV light, and have repeated the coating for enhanced tightness.
Usually, the resin cures rapidly under UV light.
However, to ensure complete curing especially in areas with limited light exposure, an additional 24-hour bake-out in an oven at 70\,\textcelsius~is recommended.

\subsection*{MSLA Printing}
As for the FDM print, we also recommend choosing materials that are capable of maintaining structural integrity at temperatures up to 50\mbox{--}70\,\textcelsius~for prolonged periods of time.
In our case we utilised an ABS-like resin.
With the Mars 4 Ultra, the printing time was about 4.5~hours.
The remaining resin on the part was washed off, and afterwards the part was fully cured using UV-light with the washing and curing station Mercury Plus 2 from ELEGOO.
The resin is also used after assembly as a sealing, see \autoref{sec:construction:assembly}.

\subsection{Scintillator Preparation}
The opaque scintillator is prepared as a mixture of three components~\cite{Buck_2019}: the liquid linear-alkylbenzene\footnote{CAS number: 67774-74-7, e.g.~from Helm AG, Sasol Limited, or BOCSCI Inc.}, which makes up 89.7\,wt.\%~(per-cent by weight), the powder 2,5-diphenyl-1,3-oxazol\footnote{CAS number: 92-71-7, e.g.~from Sigma-Aldrich/Merck KGaA or Thermo Fisher Scientific Inc.}, which makes up 0.3\,wt.\%, and small pallets of paraffin wax\footnote{CAS number: 8002-74-2, e.g.~from Thermo Fisher Scientific Inc., product number P/0600/90} with melting point of 58\,\textcelsius, which make up 10.0\,wt.\% of the mixture.
All three components are combined in a beaker and heated to 60\,\textcelsius, e.g.~on a boiler plate.
At this temperature, the wax pellets melt and blend with the linear-alkylbenzene into a transparent liquid.
The diphenyl-oxazole dissolves in the linear-alkylbenzene during the heating process, as well.
When cooling to ambient temperature, the mixture transitions from a transparent inviscid liquid at above {30\,\textcelsius} to a viscous opaque liquid at below 25\,\textcelsius~\cite{Buck_2019}.
In between, it is translucent.
This behaviour is the key feature explored in \autoref{sec:measurements:temperature}.
The liquid can be picked up by a syringe with large opening (e.g.~1\,mm diameter) for filling of the detector vessel.

\subsection{Assembly of Detector}
\label{sec:construction:assembly}
The internal fibres are cut to length and polished on one end by milling with a diamond cutter.
The fibres are then glued into the FC/PC connectors, so that the polished surface sits flush with the end of the connector, see \autoref{fig:schnitt}. For this, a two component epoxy glue is used.
After the glue is cured to a degree that the fibres are fixed in place, the connectors are screwed onto the mating sleeves and every seam is glued shut.
The mating sleeves are then glued to the vessel, while ensuring that the fibres are in the blind holes at the back of the vessel.
After the glue is cured, the lid is screwed onto the vessel, with an o-ring in between as a seal. The small hole in the centre of the lid, the main purpose of which is to let in light of an LED, is used to fill in the scintillator with a syringe. 
The FC/PC mating sleeve for the light source is milled down to the flange on one side and sealed on that side with a piece of transparent tape before screwing it onto the lid with an additional o-ring in between.
This careful sealing procedure is necessary, because the oily scintillator liquid is known to cause leakage.
After printing and assembly, the parts still need to be sealed with UV-curing resin at the joints, connectors, and the screws to ensure liquid tightness. A brush is used to apply the resin, and the same UV-light curing station as described in \autoref{sec:3Dprinting} is employed. If it turns out that the detector is not completely liquid-tight, the coating as described above can be repeated.

\section{Measurements}
\label{sec:measurements}
In this section, we describe two measurements possible with our detector, i.e.~the temperature dependent confinement of light through opacity and the measurement of cosmic elementary particles.
Both measurements require a fast light sensor and corresponding electronics or an oscilloscope.
For the temperature-dependence measurement a pulsed light source is needed in addition.
In the following, the setup we are currently using is described.

\subsection{Setup with CAEN Educational Kit}
\label{sec:CAENkit}
To read out the light from the detector the CAEN\footnote{CAEN, S.p.A. \url{https://www.caen.it}} Educational Photon Kit is used~\cite{Caccia_2013, Arosio_2014}.
The kit we are using contains a SiPM (Hamamatsu\footnote{Hamamatsu Photonics, K.K. \url{https://www.hamamatsu.com}} MPPC S13360-1350CS) assembled on a PCB inside a light-tight holder with an FC/PC fibre connector on one side and mounted to a power supply and amplification unit on the other side.
The power supply and amplification unit (SP5600) has two independent channels providing bias voltage, analog signal amplification and integrated leading edge discrimination.
In addition, a coincidence logic is implemented.
The amplified analog signals are connected to a two-channel 12-bit 250\,MS/s digitiser with an input range of $\pm$1V (DT5720A).
The digitiser is running a pulse processing firmware featuring charge integration.
Both the SP5600 and the DT5720A are connected to a computer via USB. The software CAEN HERA is used for device control and data read out.

Furthermore, the kit includes an LED driver (SP5601) with an FC/PC fibre connector, emitting about 400\,nm nanosecond light pulses and synchronous logic signals for triggering.
Alternatively to the above-mentioned SiPM holder, cables can also be attached to the SP5600.
In this way, a detector which already includes a SiPM can be read out. 
Thus, the kit allows experiments with light from the LED pulser passing through fibres and/or active material before reaching the holder with the SiPM, as well as experiments with complete external detector units.
In both cases, an oscilloscope could also be used instead of the digitiser.

\subsection{Temperature Dependent Confinement of Light}
\label{sec:measurements:temperature}
As already mentioned above (see e.g.~\autoref{fig:schnitt}), the detector assembly has been designed such that existing heating rings can be used to change the temperature of the opaque scintillator inside\footnote{These heating rings were originally added to the CAEN kit to heat up the holders containing the SiPMs to study their temperature dependence.}. 
Stainless steel nozzle heating tapes with wound resistance ribbon are used, which have an inner diameter of 40\,mm and a length of 20--25\,mm. 
If operated at 230\,V supply voltage they can reach a temperature of 400\,\textcelsius~and 8\,W/cm$^2$ heating power.
However, we limit the power by using a 36\,V 45\,W supply unit. Therefore, only temperatures up to $(65\pm 5)$\,\textcelsius~are reached.

\begin{figure}[tb]
    \centering
        \includegraphics[scale=0.35]{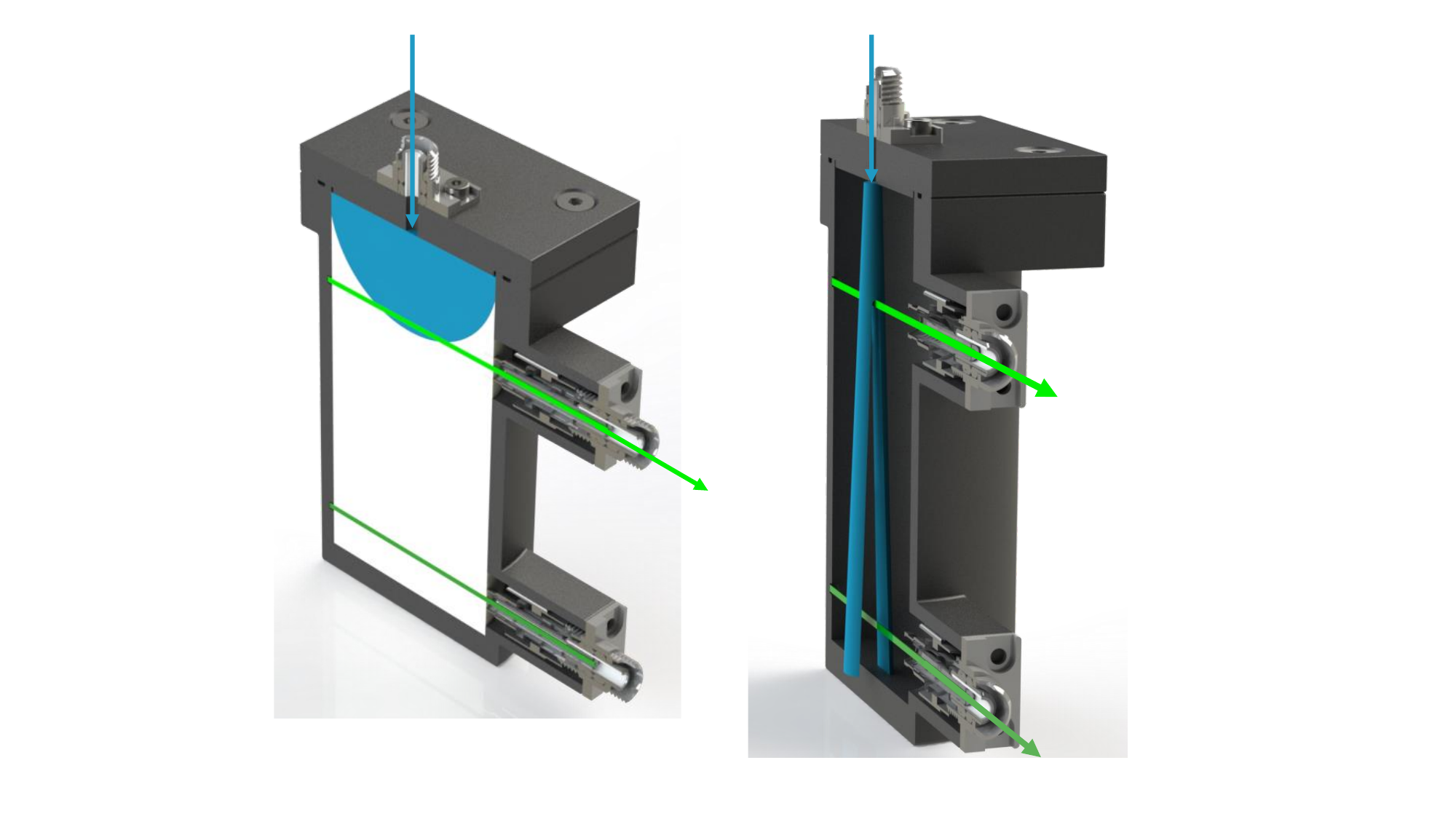}
        \caption{Artistic rendering of light propagation (blue) inside the detector using an LED as light source. Left: scintillator opaque. Right: scintillator transparent. In the configuration shown here, the lower fibre lies in the shadow of the upper fibre and thus collects much less light.}
    \label{fig:scintillatormodes}
\end{figure}

The main idea of the temperature dependent measurement is visualised in \autoref{fig:scintillatormodes}.
Light pulses created by the LED driver are guided through an optical fibre to the detector and enter at the top connector.
At room temperature, the scintillator is opaque (left image) and the light is scattered and confined in the upper part of the detector. 
When the detector is warmed up, the situation changes as the scintillator becomes transparent (right image).
Now the light can travel all the way down to the bottom of the detector.
Depending on the position of the readout fibres (the two horizontal ones in \autoref{fig:scintillatormodes}), they will see more or less light depending on the opaqueness of the scintillator.
If the lower fibre is placed in the shadow of the upper one, it will always receive much less light, as indicated in the right image of \autoref{fig:scintillatormodes}.

\begin{figure}[tb]
    \centering
        \includegraphics[scale=0.7]{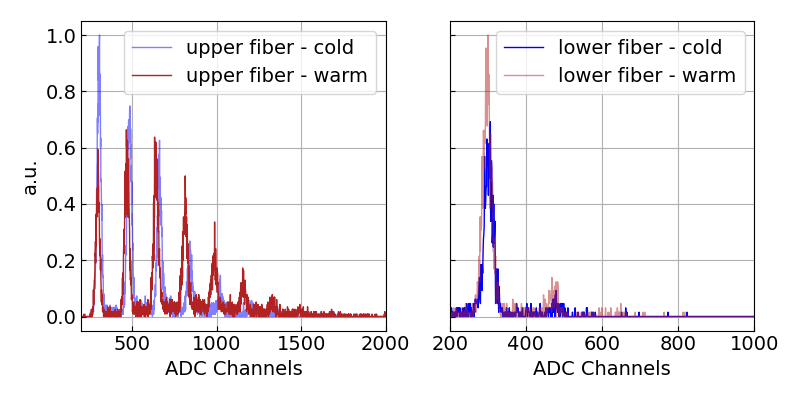}
        \caption{Spectra of the printed cell using the pulsed LED described in \autoref{sec:CAENkit}, the noise levels are omitted. The spectra are first normalised to the event count. Then the spectra of the upper fibre are normalised to the highest peak in cold conditions and the spectra of the lower fibre are normalised to the highest peak in warm conditions. The difference between the peak distances for the warm and cold spectra are due to different SiPM temperatures ($\Delta T \approx$ 2\,\textcelsius).}
    \label{fig:fingerplot}
\end{figure}

In \autoref{fig:fingerplot}, the results from four different measurement series are shown, demonstrating the above-mentioned behaviour.
For both cold and warm conditions, a spectrum is shown for the upper fibre (left) as well as for the lower fibre (right).
These spectra have been obtained using the CAEN readout electronics described in \autoref{sec:CAENkit}.
For each event, the pulse processing firmware of the digitiser performs a waveform baseline correction and, within a pre-defined gate, an integration of the sample amplitudes (in ADC channels).
In other words, omitting noise peaks centred around zero, the spectra contain integrated SiPM pulses corresponding to individual photon-equivalent signals (PE) detected.
The $n$th peak from the left thus contains all SiPM pulses corresponding to $n$ PE.
For the upper fibre (\autoref{fig:fingerplot}, left), it is clearly visible that more light is detected in the warm condition than in the cold condition, since the whole spectrum is shifted to the right, i.e.~starting from the fourth peak the red histogram contains more entries per peak than the blue one.
Thus, for the upper fibre, the red histogram has on average a higher number of PE than the blue one.
This should not be confused with a second effect showing up in this plot, namely the fact that the distance of the peaks gets smaller when the detector gets warmer.
This is due to the typical temperature dependence of the intrinsic gain of the SiPMs.
Sensors close to the SiPMs have indicated that during the whole heating procedure their temperature has changed by about {2\,\textcelsius} for the measurements presented here.
In principle, this effect could be corrected, or even avoided, but it is also an interesting and educationally valuable observation.

As expected, the lower fibre receives much less light compared to the upper fibre even when the scintillator is warmed up and therefore more transparent.
On the one hand, the light has to traverse the whole detector vessel to reach the lower fibre and, on the other hand, the lower fibre is shadowed by the upper one.
Still, the measurements show that more light is recorded in warm condition (see \autoref{fig:fingerplot}, right).
To verify that the signals recorded with the lower fibre are mostly due to light pulses from the LED and not due to dark count events or background light, a dedicated measurement without LED light has been performed. For more information and, in particular, temperature-dependent dark count rate studies using a very similar SiPM, please refer to Reference~\cite{Bauss_2016}.

\subsection{Detection of Cosmic Muons}
\label{sec:measurements:muons}
\begin{figure}
    \centering
        \includegraphics[scale=0.35]{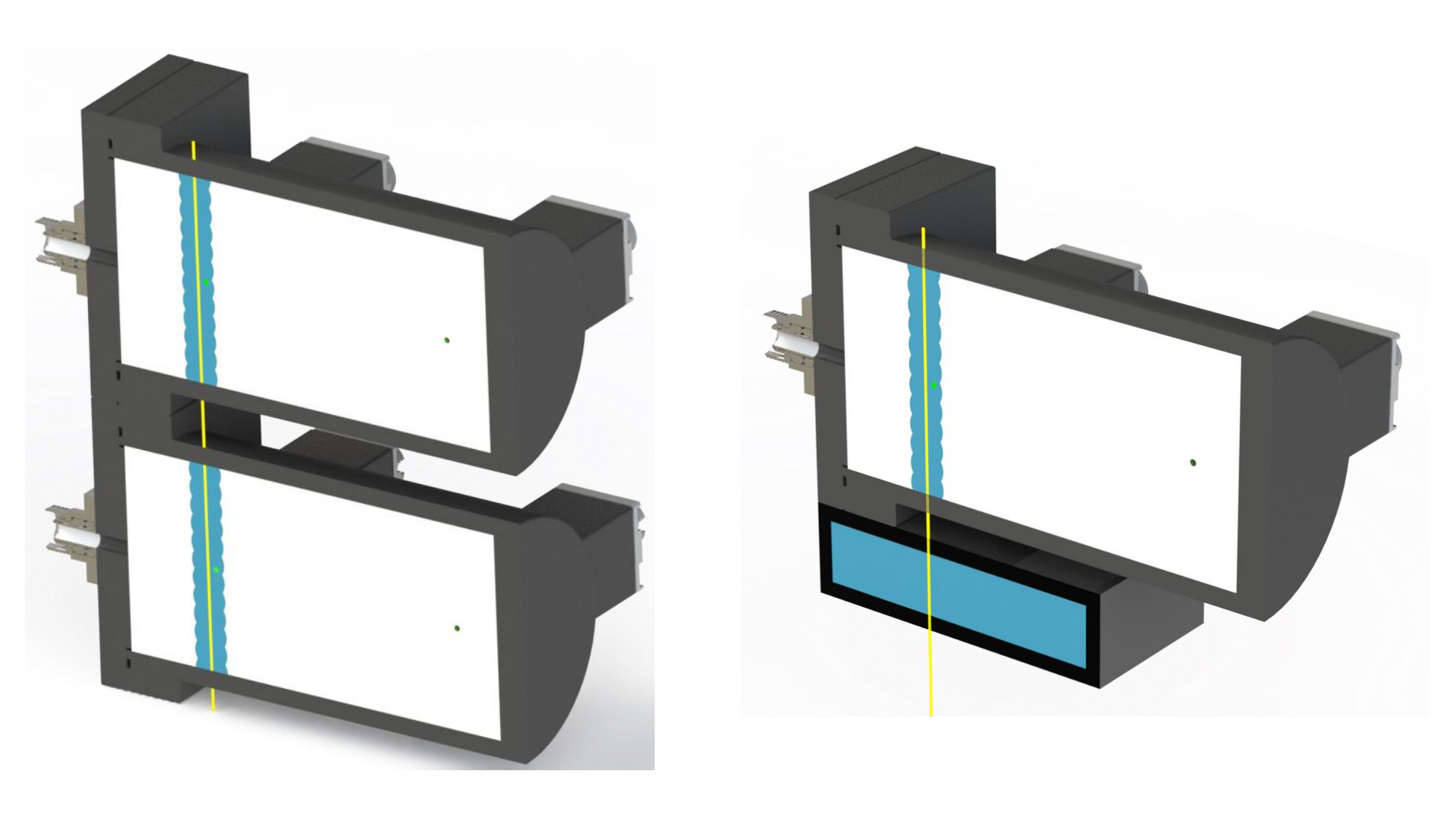}
        \caption{Artistic rendering of light production (blue) by a cosmic muon (yellow). The fibres are indicated as green dots. This setup is suitable for coincidence measurements. Left: using two opaque detectors. Right: using one opaque detector and a transparent plastic scintillator tile.}
    \label{fig:cosmic_render}
\end{figure}

For a classroom demonstration of the setup working as a particle detector, cosmic muons can be used.
While crossing the detector, they deposit part of their energy in the active medium, i.e.~in the scintillator.
The scintillation light created can be read out by the same SiPM and electronics described above.
Compared to the signals obtained when using the LED pulser, most of the signals originating from cosmic muons should be larger, but at a much lower rate.
In principle, one detector is sufficient, however, to reliably distinguish cosmic muons from radioactive background it is better to set up two detectors on top of each other to perform a coincidence measurement.
This is illustrated in \autoref{fig:cosmic_render}.
Requiring that both detectors show a signal within a time window of 100\,ns, for instance, largely reduces the probability for background signals.

\begin{figure}[tb]
    \centering
        \includegraphics[scale=0.6]{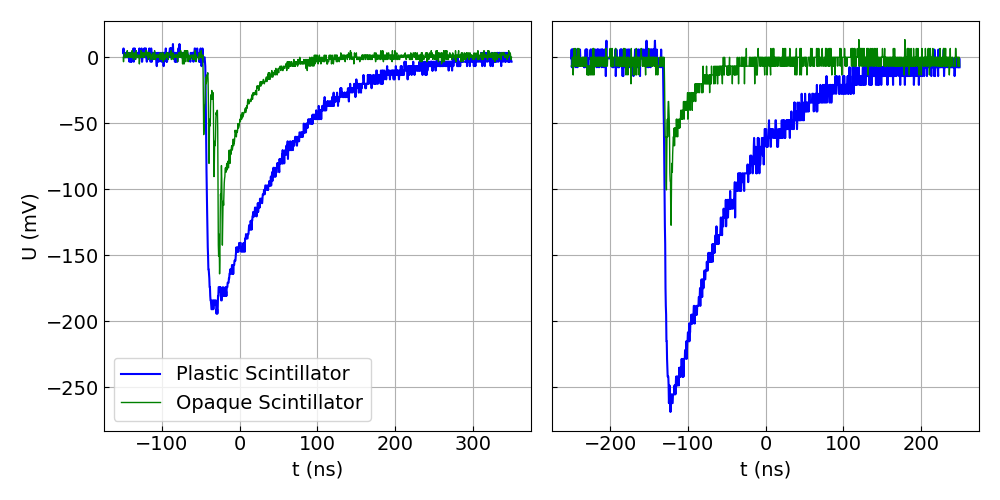}
        \caption{Two examples (left and right) of a cosmic muon signal. The respective muon was measured at the same time in both the plastic scintillator tile from CAEN (blue) and the 3D-printed cell filled with opaque scintillator (green). This data was recorded with an oscilloscope (2.5\,GS/s sampling rate, 1\,GHz analog bandwidth). While the signal from the tile shows the usual shape, the signal from the cell exhibits many overlaying signals from individual photons with a time delay of approx.~7\,ns. This is caused by the random paths traveled by the light due to scattering in the opaque medium.}
    \label{fig:cosmic_plot}
\end{figure}

Instead of a second opaque scintillator cell, a plastic scintillator can also be used for the coincidence measurement.
The CAEN kit described above can be extended by a 47\,mm~$\times$~47\,mm area plastic scintillator tile (SP5608), with a thickness of 10\;mm, which is coupled to a SiPM (Hamamatsu MPPC S13360-6050CS). 
\autoref{fig:cosmic_plot} presents two examples of a cosmic muon event recorded with an oscilloscope connected to the power supply and amplification unit of the CAEN kit.
In blue colour, the signal from the plastic scintillator tile is shown and in green colour the coincident signal from the 3D-printed opaque scintillator cell is overlaid.
Notably, the signal shape of the latter exhibits a substructure, which is typical for individual photons being captured by the fibre and is also seen in other opaque detectors~\cite{LiquidO_2024,pin:tel-03149593}.
The spatial resolution of the measurement of cosmic muons can be improved by densely arranging several fibres in a grid, as done in a recent detector prototype~\cite{SussexCube}.

\section{Conclusion}
\label{sec:conclusion}
In this article, we described the construction of a small particle physics detector using state of the art scintillation detector technology for demonstration in educational context.
Our setup can be constructed with relatively moderate effort, given that a home-level 3D-printer is available.
In addition, a photosensor (e.g.~a SiPM) is required, as well as some equipment for the signal readout.
Ideally, dedicated readout electronics including an amplifier are used, however, even with an oscilloscope some measurements can be done.
The setup allows to understand modern detector technology for particle physics in an engaging hands-on experience and is suitable to detect elementary particles.

\section*{Acknowledgements}
We thank Cloé Girard-Carillo for rendering \autoref{fig:opaque_topology} for us.
This work has been supported by the Cluster of Excellence ``Precision Physics, Fundamental Interactions, and Structure of Matter'' (PRISMA$^{+}$ EXC 2118/1) funded by the German Research Foundation (DFG) within the German Excellence Strategy (Project ID 390831469).

\appendix
\section{}
\label{sec:appendix}
\begin{figure}[tb]
    \centering
        \includegraphics[width=\linewidth]{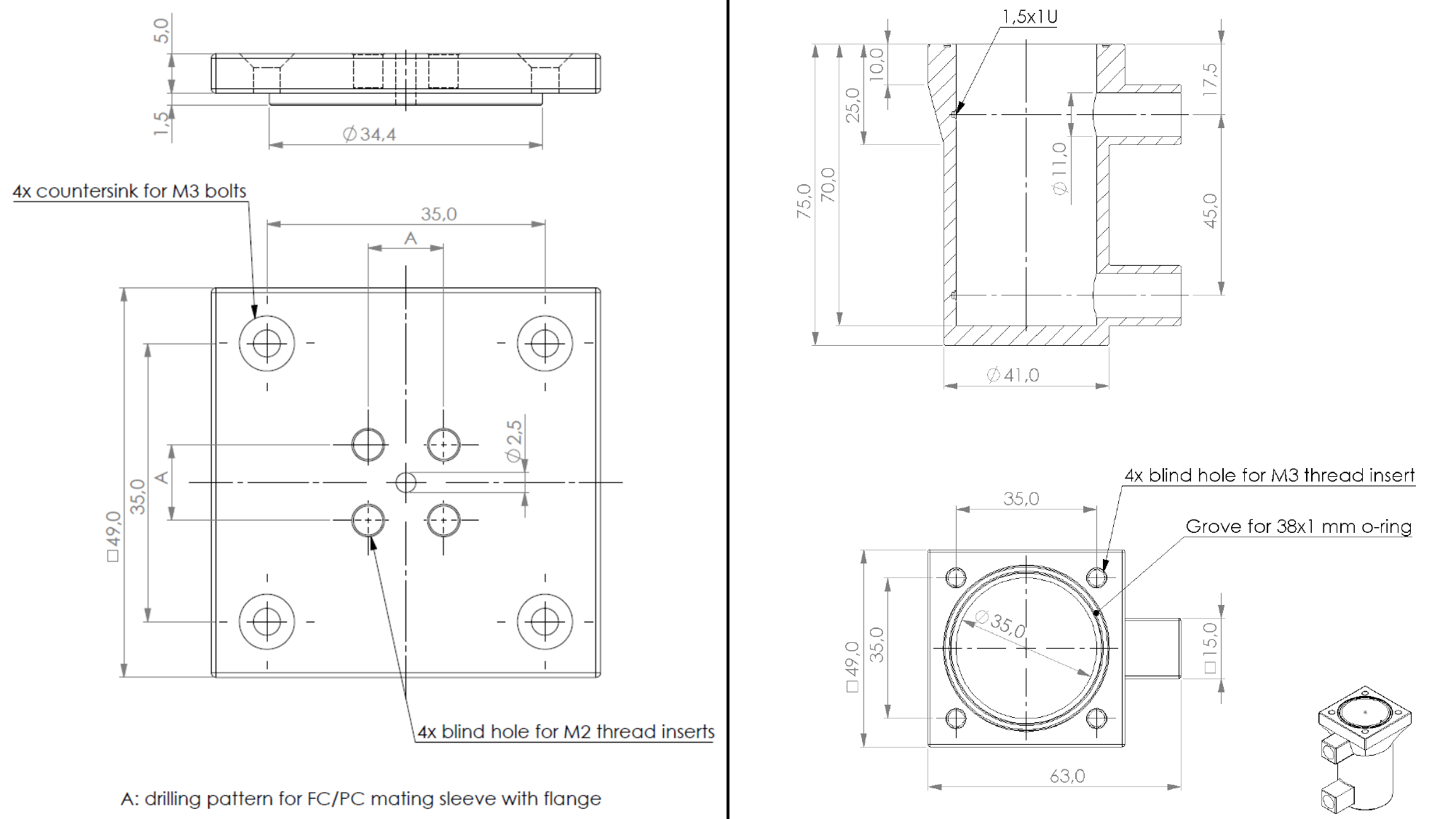}
        \caption{Technical drawings of the lid (left) and the vessel (right)~\cite{zenodo_files}. All edges chamfered 0.2\,mm.}
        \label{fig:technicaldrawings}
\end{figure}

Technical drawings of the lid and vessel are shown in \autoref{fig:technicaldrawings}.
CAD drawings and STL data files of the lid and vessel can be found at Reference~\cite{zenodo_files}.

\section*{References}
\bibliography{References}   

\end{document}